# A review of the pentaquark $\Theta^+$ properties

A. R. Haghpayma[†]

*Department of Physics, Ferdowsi University of Mashhad*

*Mashhad, Iran*

## Abstract

Although the $\Theta^+$ has been listed as a three star resonance in the 2004 PDG, its existence is still not completely established, Whether the $\Theta^+$ exist or not, but it is still of interest to see what QCD has to say on the subject. for example, we should know why the $\Theta^+$ width is extremely narrow. in this paper i review briefly the pentaquark $\Theta^+$ properties.

### I. INTRODUCTION

The year 2003 will be remembered as a renaissance of hadron spectroscopy at the earlys of that year ( LEPS ) collaboration, T. Nakano et al.[1] reported the first evidence of a sharp resonance $Z^+$ renamed to $\Theta^+$ at $M_{\Theta} \simeq 1.54 \pm 0.01$ Gev with a width smaller than $\Gamma_{\Theta} < 25 MeV$

The experiment performed at the Spring-8 facility in japan and this particle was identified in the $K^+N$ invariant mass spectrum in the photo-production reaction $\gamma n \to K^- + \Theta^+$, which was induced by a Spring-8 tagged photon beam of energy up to 2.4 Gev.

The existence of $\Theta^+$ was soon confirmed by various groups in several photo—nuclear reactions[2] including V.V Barmin et al.[3] ITEP( DIANA )[4] JLAB¹( CLAS )[5] and ELSA( SAPHIR ).

Perhaps the simplest data coms from SAPHIR detector[6] at ELSA where the $\Theta^+$ is photo— produced of a simple target. The final state contains $nK^+K^0_s$ and the relevant system is identified in the missing mass spectrum of the $K^0_s$. The $K^0_s$ is reconstructed from its two $\pi$ decay, preferentially in the forward direction. these authors conclude the $\Theta^+$ is an isoscalar due to the absence of a $\Theta^{++}$ in the $\gamma p \to pK^+K^-$ channel.

After the jefferson lab confirmation, it was observed in several different experiences, with a mass of $1540 \pm 10$ Mev and a decay width of $15 \pm 15$ Mev.

Since 2003, january there have been several reports of exotics:[8]

Because of the observation of such states in various reaction channels, the existence of pentaquark baryons now becoms widely accepted,.:

Such states are believed to belong to a multiplet of states where the possible observability of the other members has to be worked out

This discovery has triggered an intense experimental and. theoretical activity to understand the structure of the state

With the conventional constituent quark model, the conservation rules guarantee that it has a strangeness S=1, baryon number B=1 and charge Q=1, thus the hypercharge is $Y = B + S = 2$ and the third component of isospin is $I = 0$. no corresponding $pK^+$ ( I=1 ) state is observed at the same mass, due to absence of a $\Theta^{++}$ in the $\gamma p \to pK^+K^-$ channel and thus the isospin of $\Theta^+$ is the same I =0 and it also seems important that no S=1 baryon states has been observed below the NK threshold, and this state seems to be the ground state

We have two decays $\Lambda$ ( 1540 ) $\to KN$ and $\Lambda$ ( 1600 ) $\to KN$ above the threshold but both decays need $q\bar{q}$ pair production from vaccum, but we have for $\Theta^+$ decay: $\Theta^+ \to K^+N$ and it seems that no need $q\bar{q}$ pair production if $\Theta^+$ is not a more complicated object

All known baryons with B = 1 carry negative or zero strangeness. a baryon with strangeness S = 1, it should contain at least one $\bar{s}$, can not consist of three quarks, but must contain at least four quarks and an antiquark; in other words, must be a pentaquark or still more complicated object. now its called $\Theta^+$ pentaquark in literature.

From the charge and the strangeness. $u^2 d^2 \bar{s}$ is a possibility as the content of $\Theta^+$, which called the minimum quark content, such state is exotic; in general states with the $\bar{q}$ having different flavour, than the other four quarks and their quantum numbers cannot be defined by 3 quarks alone are called exotics thus we have an exotic $\Theta^+$

The discovery of $\Theta^+$ was followed by the discovery of yet another exotic baryon, $\Xi^{--}$, found by the NA40 group at CERN[7] with M=1862± 0.02 Mev, the particle ($\bar{s}d^2u^2$) is another manifestly exotic baryon whose decay at $\Xi^- \pi^-$ has been observed at the mass M =1.862 Gev with a width $\Gamma$<18 Mev and the uuddc pentaquark was $D^*P( 3100 )$ observed at H1[8], its mass 3099 ± 3 ± 5 and width compatible with experimental resolution, decaying to $D^* P$ which search for but not seen by Zeus[8].

Another report from WA89 colaboration shows no evidence for $\Xi$ ( 1860 ) in $\Sigma^-$ – nucleus collisions[9], however higher statistics experiments are required to firmly establish the observed states[10].

The possibility and the interest for S =1 baryons (or Z baryons) has been recorded for many years by the PDG up to 1986. but subsequently it was dropped because of lack of clear evidence for their existence.

The efforts to search for pentaquark baryons until 1980's were summarized in Ref [11].

However this exotic baryon with such a low mass and so narrow a width impose a big challenge to hadron theories and its discovery shall be one of the most important events in hadron physics.

If it is really a pentaquark state it will be the first multi-quark states people found. theoretical interest in exotic baryons has continued both for heavy and light quarks[12]. the first prediction of the mass of $\Theta^+$ is $M_z$= 1530 Mev by Mpreszalowicz at 1987 and the first prediction of width of $\Theta^+$ is $\Gamma_z < 15$ Mev by M.Polyakov,

---

[†]e-mail: haghpeima@wali.um.ac.ir

¹ Although the JLab new experiment[21] searched for pentaquarks in the same channel at a level at least one order of magnitude better-than the previous published result and found no pentaquarks., in general, the negative reports which involve higher energies and statistics or reactions different from those that produced positive results may not directly contradict them.

D.Diakonov , V . Petrov. at 1997.

The averaged mass value $M_\Theta =$ 1536 .2 $\pm$ 2.6 Mev and $\Gamma_\Theta =$ 12 $\pm$ 9 $\pm$ 3 Mev and the world average is $M_\Theta \simeq$ 1538 Mev

The mass and the width of $\Theta^+$ and other exotic pentaquark baryons has predicted by several hadron models . its width ($\Gamma < 10$ Mev ) is exceptionally narrow as for a hadron resonance located at 110 Mev a bove the NK threshold usually refered to narrow width puzzle.

There is no direct measurement of its spin S and Isospin I and its angular momentum J and parity P are different in various theoretical works, however most of them postulated its angular momentum J to be J = 1/2 but the possibility of J = 3/2 and S=1/2 and P= + is rather plausible.

If some of theoretical models is correct, there should be new pentaquark states waiting for discovery. are these new states existe the answer is experiment but in the experiments we see a varietyof mass for $\Theta^+$ , are they same particle and the differences are due to experimental errors.

Where else to look for $\Theta^+$ pentaquark production.

1. Nucleon- nucleon collisions:
2. Photon - nucleon collisions:
3. Pion - nucleon collisions:

Measurement of parity is crucial to test theories and there are many suggestions on detecting the parity. .another test is measurement of the isospin and spin of pentaquark states .

If $\Theta$ confirmed and established,a new landscape of multi-quark hadrons is emerging from the horizon. we must answer what's the underlying dynamics leading to its low mass, narrow width and special prioduction mechanism ?

Do other multiquark hadrons exist ? 4q ,6q ,7q ,… , Nq, is there an upper limit for N ? study of these issues will deepen our understandig of the low- energy sector of QCD.

QCD as the underlying theory of hadron theories predicts beyond 3q or $q\bar{q}$ also multiquark states - quark gluon hybrids - glueballs and so on but it is unperturbative at low - energy , thus we need to understand the underlying dynamics of these states and concepts such as confined quarks and gluones or fundamental concepts such as spin and mass of a confined quark and a free quark or the width of decay of a hadron.

Although the $\Theta^+$ has been listed as a three star resonance in the 2004 PDG , its existence is still not completely established:

Whether the $\Theta^+$ exist or not, but it is still of interest to see what QCD has to say on the subject.

## II. NARROW WIDTH PUZZLE

The pentaquark width ( $\Gamma < 10$ ) is exceptionally narrow as for a hadron resonanc located at 110 Mev a bove the NK threshold both $\Theta^+$ and $\Xi^{--}$ are very narrow states. $\Theta^+$ is so narrow that most of the experimental results show only an upper bound around 20 Mev or from the recent KN scattering is less than several Mev [13].

While the width of conventional exited hadrons always are around one hundred Mev or even bigger , if they lie 100 Mev a bove threshold and decay through S -wave or P -wave.

For comparison the S= 1 hyperon $\Lambda$ ( 1520 ) $D_{03}$ state $J^P = 3/2^-$ and in the same mass region as the $\Theta^+$ , has dominant two - body decay D - wave with final states NK , with a smaller phase space and higher partial wave, and its width is 15.6 - Mev [14] while $\Theta^+$

P -wave with a total width less than several Mev, corresponding to negative or positive parity respectively $\Lambda$ (1520) decay to KN through D - wave and its width is 7 Mev also $\Delta$ ( 1600 ) decay to K N through P - wave and its width is 100 Mev this two decays need $q\bar{q}$ creation and $\Theta^+$ is in the same phase space but its width is smaller than 10 Mev ,there is a puzzle.

The question is the origin of narrow width of the pentaquark which is the most peculiar feature of this new resonance. in other words is there mysterious selection rule which is absent from the conventional hadron intraction ?

Since there is no known selection rule from symmetry to make the width naturally small, the narrow width should have dinamical origin. can low - energy QCD describe the underlying dinamical forces between quarks and gluons in such states and generate their mass and width correctly ?

There have been several attempts to explain the narrow width from combinational suppression from the spin - flavor and color factors or from the special spatial structure due to diquark correlations and from the theories which describe the behavor of quarks and gluons like chiral soliton and instanton liquid models.

Several attempts performed on confinement. however, no difinite conclusion has been reached yet.

A conventional dynamical mechanism for the long life time of the pentaquark state is treating its hadronic decay on the basis of the constituent quark model which is supported by QCD as underlying theory and molecular dynamics. this mechanism suppose that the constituent quarks are well mixed in the colors space inside the pentaquark and rearrangement of their colors, flavors, spins and spatial positions into two color- white clusters i. e. the Nucleon N and the Kaon K takes a long time and thus gives a narrow width for the pentaquark decay.

This rearrangement or regrouping is goverend by the strong interactions among quarks and is not simply related to the distance in color - flavor - space between the initial and final states.

This method is called color molecular dynamics (CMD). in this method for example the wave function of a single quark as a constituent one is parametrized by a gaussion wave pocket in coordinate space and by a color coherent state in the $SU_C(3)$ and so on. then by using of Hamiltonian commonly used in the standard constituent quark models [15]. the Time - dependent dynamics of the multiquark systems in their clustres and decays are explained .

Because of the importance of strong forces in QCD , color and spatial coordinates is more interesting and spin - flavours and antisymmetrization sometimes are neglected, however their effects on the width and mass of a state are not important in QCD.

There have been several attempts to explain the narrow width of the $\Theta^+$ pentaquark [17].

For the $\Theta^+$ the most efficient decay mechanism is for the 5 quarks to regroup with each other into a three - quark baryon and a meson, that is in contrust to the $3P_0$ decay models of the ordinary hadrons [18].

Carlson et al. constructed a spacial pentaquark wave function which is totally symmetric in the flavor- spin part and anti- symmetric in the color - orbital part [19] ,with this wave function they found the overlap probability between the pentaquark and the nucleon kaon system is 5/94.

Taking into account of orbital wave function in JW s diquark model further reduces the overlap probability to 5/594.

The small overlap probability might be responsible for the narrow width of pentaquark.

Another approach to this puzzle is a $SUSY^{[29,30]}$ based model in which we use a broken dynamical supersymmetry between an antiquark and a diquark by replacing two antiquarks in an antibaryon by two diquarks to form a pentaquark and relating their mass to each other.

Using this technique , we find that the mass of an exotic pentaquark with strangeness $S = 1$ is at least 200 Mev larger than that of the reported $\Theta^+$ pentaquark.

Furthermore, there is no reason for the pentaquark to be narrow, and the pentaquark should have such a broad width as to make it hard to observe as a resonance, even in the case with orbital angular momentum $\ell = 0$ .

It seems if we use another unknown dynamical interactions beetween quarks in diquark after replacing it , we will be able to decrease the mass and decay width of the pentaquark , if this replacement ocure in color space only, we can use the QCD based behaviour of spin, flavors and antisymmetrization of diquarks to describe such dynamical interactions.

A few years a go similar ideas were used to predict the masses of exotic mesons and baryons but not for the properties of pentaquarks [16].

The standard constituent quark model is the next approach to this puzzle, at the symmetry limit the selection rules are exact and the narrow width of the pentaquarks come from the symmetry breaking.

Buccella and sorba suggested [20] that the four quarks are in the L=1 state and the anti- quark is in the S - wave state inside the $\Theta^+$ and $\Xi^{--}$ .

There are four anti - symmetric four quark $SU_{fs}(6)$ wave functions that are

$[4]_{fs}$  $[31]_{fs}$  $[22]_{fs}$  $[211]_{fs}$

in the diquark model (JW) the four quarks are not completly anti-symmetric, thus the thired one correspond to this model.

The narrow width of $\Theta^+$ pentaquark may favor the last two wave functions[31].

The reason is as follows: when the anti-quark picks up a quark to form a meson, two of the other three quarks remain in the $SU_{fs}(6)$ totally symmetric representation $[3]_{fs}$ for the nucleon octet.

If $SU_{fs}(6)$ symmetry is exact, the $\overline{10}$ pentaquarks will not decay at all, this selection rule is exact in the symmetry limit, the narrow width of the $\Theta^+$ and $\Xi^{--}$ pentaquarks come from the $SU(3)_f$ symmetry breaking.

In order to refine our understanding of quark dynamics at low energy where it is not perturbative we review some general features of the dynamics of a $K^+N$ resonance.,

$\Theta^+$ lies about 100 Mev above $K^+N$ threshold at a center of mass momentum $K = 270$ Mev, the characteristic parameter $KR$ is a both 6.4 if we use a typical Range $R = 1$ F for this interaction. assuming isospin zero with $KR \simeq 1.4$ only the S or P -wave is likely and the spin $S = 1/2$ is pleasible.

QCD features are:
1- because of low center of mass momentum and no other hadronic channels coupling to $K^+N$ below $K\Delta$ threshold at 1725 Mev we are at nonrelativistic region.
2- $\Theta^+$ is an exotic particle and in its scattering to $K^+N$ there is no quark-antiquark annihilation graphs, thus we no have confined states that couple by $q\bar{q}$ anihilation.
3- the wave function of $K^+$ and N or final states differ from the $\Theta^+$ in space, color, and spin.

Acording to this features of QCD in the region of $\Theta^+$ we lead to a nonrelativistic potential scattering description of it, but this description cannot reproduce its mass and width semeaultaneously correctly the reason is as follows:

The resonance are related through the range and depth of the potential, for a simple attractive potential of Range 1 F, the width of a P-wave resonance 100 Mev above threshold is above 175 Mev and for a width of order 10 Mev our range must be 0.05 F but this range of a potential brings in a high energy scale, far from th$\Theta^+$ P-wave resonance.

Thus one can choos the potential Range to be 1 F and decrease the mass of $\Theta^+$ by some additional dynamics beyond nonrelativistic potential scattering, for example: hyperfine interactions such as flavour-spin and color-spin interactions between quarks inside the $\Theta^+$ and confinement effects.

Although this is our understanding of quark dynamics at low energy there is various attempts to refine it.

We can produce the mass of $\Theta$ by decoupling of decay modes through mass matrix diagonalization between pentaquark state for example two degenerate pentaquark octet and antidecouplet. (JW).

However in JW's model $\Theta^+$ is not the lightest pentaquark due to the ideal mixing between the octet and anti-decuplet the ideal mixing will split the spectrum and produce two nucleon-like states ($N_s(1710)$, $N_1(1440)$) wilczek identified $N_1$ as the well-known Ropper resonance $N(1440)$ which is a very broad four-star resonance[32] with a width around (250 to 450 Mev).

However, it will be very demanding to explain Ropers larg decay width and $\Theta^+$'s extremely narrow width simultanously[33] which excludes $N_1(1440)$ as a pentaquark state, $N_s(1710)$ has a large branching ratio into $\pi\Delta$ channel, and it should be excluded as a pure antidecouplet state.

This is because, within $SU(2)$ symmetry, antidecouplet does not couple to decouplet and meson octet, therefor mixing with other multiplets is required if one want to identify $N_s(1710)$ as a pentaquark Crypto-exotic[22-28] states.

## References

[1] T. Nakano et al., Phys. Rev. Lett. **91**, 012002 (2003).

[2] S.Stepanyan et al., CLAS Collaboration, Phys. Rev. Lett. **91**, 252001 (2003), hep-ex/0307018; V. V. Barmin et al., DIANA Collaboration, Phys. Atom. Nucl. **66**, 1715 (2003) [Yad. Phys. **66**, 1763 (2003)], hep-ex/0304040; J. Barth et al., SAPHIR Collaboration, Phys. Lett. **B572**, 127 (2003), hep-ex/0307083; A. E. Asratyan et al., hep-ex/0309042, to be published in Phys. Atom. Nucl.; V. Kubarovsky et al., CLAS Collaboration, Phys. Rev. Lett. **92** (2004) 032001; A. Airapetian et al., HERMES Collaboration, hep-ex/0312044; A. Aleev et al., SDV Collaboration, hep-ex/0401024.

[3] V.V. Barmin et al., Phys. Atom. Nucl. **66**, 1715 (2003); Yad. Fiz. **66**, 1763 (2003).

[4] DIANA collaboration, V.V. Barmin, et al., Phys. Atom. Nucl **66** (2003), 1715.

[5] CLAS collaboration, S. Stepanyan, et al., Phys. Rev. Lett **91** (2003), 252001.

[6] J. Barth et al, [SAPHIR Collaboration] arXiv:hep-ex/0307083.
SAPHIR Collaboration, J. Barth et al., Phys. Lett. B **572** (2003), 127.

[7] NA49 Collaboration, C. Alt et al., Phys. Rev. Lett. **92** (2004), 042003.
C. Alt et al., NA49 Collaboration, Phys. Rev. Lett. **92** 042003 (2004), hep-ex/0310014.

[8] H1 Collaboration, A. Aktas et al., hep-ex/0403017.

[9] WA89 Collaboration, M. I. Adamovich et al., hep-ex/0405042.

[10] J. Pochodzalla, hep-ex/0406077.

[11] E. Golowich, Phys. Rev. D **4**, 262 (1971).
Particle Data Group, M. Aguilar-Benitez et al., Phys. Lett. B **170**, 1 (1986).
H. Gao and B.-Q. Ma, Mod. Phys. Lett. A **14**, 2313 (1999).

[12] H. Högaasen and P. Sorba, Nucl. Phys. B **145**, 119 (1978);
M. de Crombrugghe, H. Högaasen and P. Sorba, Nucl. Phys. B **156**, 347 (1979).
A.V. Manohar, Nucl. Phys. B **248**, 19 (1984).
M. Chemtob, Nucl. Phys. B **256**, 600 (1985).
M. Praszalowicz, in Skyrmions and Anomalies (M. Jezabek and M. Praszalowicz, eds.), World Sci-entific (1987), 112-131; M. Praszalowicz, Phys. Lett. B **575**

[13] S. Nussinov, hep-ph/0307357; R. A. Arndt, I.I.Strakovsky, R. L. Workman, Phys. Rev. C **68**, 042201(R) (2003); J.Haidenbauer and G. Krein, hep-ph/0309243; R. N. Cahn and G. H. Trilling, hep-h/0311245.

[14] ParticleDataGroup, Phys.Rev.D **66**, 010001(2002)

[15] See, e.g. M.Oka and K.Yazaki, inuarks and Nuclei, ed. W.Weise (World Scientific, Singapore, 1984), p.489.

[16] D.B. Lichtenberg, R. Roncaglia, and E. Predazzi, J. Phys. G Nucl. Part.Phys. **23**, 865 (1997).

[17] M. karliner and H.J. Lipkin, hep-ph/0410072.

[18] A. hang et al., hep-ph/0403210.

[19] C. E. Carlson et al, hep-ph/0312325.

[20] F.Buccella and P.Sorba, hep-ph/0401083

[21] CLAS collaboration, B. Mckinnon, et al ,arXiv:hep-ex/0603028 V1 14 Mar 2006

[22] T.Hirose, K.Kanai, S.Kitamura, and T.Kobayashi, Nuovo Cim. **50A**, 120 (1979)
C.Fukunaga, R.Hamatsu, T.Hirose, W. Kitamura, and T. Yamagata, Nuovo Cim. **58A**, 199 (1980).

[23] V.M.Karnaukhov, V.I.Moroz, C.Coca, and A.Mihul, Phys. Lett. B **281**, 148 (1992).

[24] A.V.Aref'ev, et al., Yad. Fiz. **51**, 414 (1990) [Sov. J. Nucl. Phys. **51**, 264 (1990)].

[25] J.Amirzadeh, et al., Phys. Lett. **89B**, 125 (1979).

[26] B.M.Abramov, et al., Yad. Fiz. **53**, 179 (1991) [Sov. J. Nucl. Phys. **53**, 114 (1991)].

[27] D.Aston, et al., Phys. Rev. D **32**, 2270 (1985).

[28] L.G.Landsberg, Phys.Rep. **320**, 223 (1999); SPHINX Collaboration, Yu. M. Antipov, et al., Yad. Fiz. **65**, 2131 (2002) [Phys. Atom. Nucl. **65**, 2070 (2002)].

[29] S.Catto and F.Gürsey, Nuovo Cimento **86**, 201 (1985).

[30] M.Anselmino et al., Rev. Mod. Phys. **65** 1199 (1993)

[31] F.Buccella and P.Sorba, hep-ph/0401083

[32] R. Jaffe and F. Wilczek, Phys. Rev. Lett. **91**, 232003 (2003).

[33] F. E. Close and J. J. Dudek, Phys. Lett. B **586**, 75 (2004).
S. H. Lee, H. Kim, and Y. Oh, hep-ph/0402135.